\documentstyle[twocolumn,floats,aps,prl,epsf]{revtex}	

\begin{document}
\title{$^{29}$Si NMR and Hidden Order in URu$_{2}$Si$_{2}$}
\author{O. O. Bernal, C. Rodrigues, and A. Martinez}
\address{Physics Department, California State University,
         Los Angeles, CA 90032}
\author{H. G. Lukefahr}
\address{Physics Department, Whittier College, Whittier, CA 90608}
\author{D. E. MacLaughlin}
\address{Physics Department, University of California, Riverside, CA 92521}
\author{A. A. Menovsky, and J. A. Mydosh}
\address{Kamerlingh Onnes Laboratory, Leiden University, The Netherlands}

\author{\small(\today)}						

\address{\parbox{14cm}{\bigskip\rm\small			
We present new $^{29}$Si~NMR spectra in URu$_{2}$Si$_{2}$
for varying temperature $T$, and external field ${\bf H}$.
On lowering $T$, the systematics of the low-field lineshape 
and width reveal an extra component $\lambda$ to the linewidth  
below $T_N\sim{17}$~K not observed previously.
We find that $\lambda$ is magnetic-field independent and 
dominates the low-field lineshape for all 
orientations of ${\bf H}$ with respect to the tetragonal 
$c$~axis.
The behavior of $\lambda$ indicates a direct relationship 
between the $^{29}$Si-spin and the transition at $T_N$, 
but it is inconsistent with a coupling of the nuclei to 
static antiferromagnetic order/disorder of the U-spin 
magnetization.
This leads us to conjecture that $\lambda$ is due to a 
coupling of $^{29}$Si to the system's hidden-order parameter.
A possible coupling mechanism involving charge degrees of 
freedom and indirect nuclear spin/spin interactions is 
proposed.
We also propose further experiments to test for the existence
of this coupling mechanism.
\\[6pt] PACS numbers: 71.27.+a, 75.30.Mb, 76.60.Cq., 76.60.Jx.
}
}	
\maketitle



\newpage
URu$_{2}$Si$_{2}$ posesses an unusual coexistence of small-moment magnetic order
($\mu\sim{0.02}\mu_{B}$/U, $T_{N}\sim{17}$~K)
and unconventional superconductivity ($T_{c}\sim{1.2}$~K)~\cite{first}.
Understanding the magnetic and superconducting behavior
of this material has proven to be very challenging and has generated
a substantial amount of work over the years (see e.g., Refs.~\cite{first,neutrons,example,theory,experiment,nqr,mydosh2k,x,theory2k,amitsuka,kohori,kohori2k,bernaljapan,Mac}
and references therein). 
The transition at $T_N$ is still puzzling.
Although neutron diffraction (ND) experiments indicate antiferromagnetic
order~\cite{first,neutrons}, the ordered-moment size is too small
to account for macroscopic effects in the thermodynamic 
quantities~\cite{example}.
Thus, many more studies have appeared with the goal of elucidating the 
nature of the true order parameter for this transition 
(see e.g., Refs.~\cite{theory,experiment,nqr,mydosh2k,x,theory2k,amitsuka}).
Quadrupolar order has been suggested~\cite{theory};
some experiments seem consistent with this picture~\cite{experiment}, 
while others are less telling~\cite{nqr}.
Experiments are consistent with two distinct energy scales in the 
system~\cite{theory} indicating primary and secondary order parameters.
A coupling or switching between the two parameters is also apparent from 
phenomenological arguments (see~\cite{mydosh2k,x} for recent discussions).
In all, several theories involving exotic microscopic 
mechanisms have been formulated and are still a matter of controversy
and debate~\cite{theory2k,amitsuka}. 

Part of the problem can be traced back to the experimental 
characterization of the magnetic state at low external fields.
This is because the time scales from different 
techniques used to sample the low frequency ($\lesssim{100}$~MHz) 
magnetic response (i.e., magnetization, NMR, $\mu$SR)
do not overlap with those from techniques used to probe the spin 
dynamics (i.e., ND).
In particular, ND measures a ``static'' ordered moment of about 
0.02--0.03$\mu_B$~\cite{neutrons} along the $c$~axis,
which would necessarily split the $^{29}$Si~NMR line by 
$\sim$80--100~Gauss (external field collinear with the spontaneous 
moment).
This splitting is not observed~\cite{kohori,kohori2k,bernaljapan},
and as a result, it has been argued that the tiny moment 
must be fluctuating at time scales that render it invisible to NMR
or even $\mu$SR probes~\cite{kohori2k} (See, however, Ref.~\cite{Mac}).
At first, this argument seems to suggest that NMR is unable
to shed light on the nature of the phase transition.
However, since NMR can sample electronic effects both directly 
and indirectly~\cite{nmr}, it should also be expected to reveal other 
significant information about the ordered state, even if a
transition to static magnetic order at $T_N$ is not established by
NMR measurements.
Accordingly, the absence of line splitting might offer a chance to 
better characterize the transition at low fields without the added 
complication of static magnetic-order effects on the NMR lineshape.
Of particular interest, for example, would be the possibility of 
observing effects due to a charge-related order parameter.

In this Letter we report a new $^{29}$Si~NMR study at low 
field strengths (below 6~T)\@.
We find an unambiguous, field-independent isotropic component of 
linewidth~$\lambda$ 
which increases below $T_N$ to about 11.5(1.5)~G, average value, 
at 4.2~K~\cite{comment}.
This component is static and measures a distribution of local 
effective fields at the $^{29}$Si~sites.
Its temperature dependence is that of a mean-field order parameter.
We argue that $\lambda$ is unrelated to the static magnetization
of the sample and that, instead, it is due to a coupling of the 
$^{29}$Si~nuclei to the ``hidden order''~\cite{mydosh2k} in this 
system.
We propose a coupling mechanism based on charge degrees of freedom
and nuclear spin/spin interactions. 
We also propose further experiments to test for the nature of this 
coupling.

The sample used was a fine powder (particle size $\alt$~50~$\mu$m)
embedded in Stycast 1266 epoxy and oriented in a field of 9.4~T\@.
An alignment factor of order 90--95\% was estimated for the $c$~axis
orientation
by measuring the magnetic susceptibility under both transverse 
and longitudinal external fields, and comparing it
with that of a single crystal under similar conditions.
The alignment method leaves a random distribution of ($a$,$b$) 
basal-plane orientations (tetragonal structure).
The $^{29}$Si spectral parameters were measured as functions of 
temperature and applied field for different field directions with 
respect to the $c$~axis.

\begin{figure}[h]
\centerline{\epsfxsize=2.5in \epsfbox{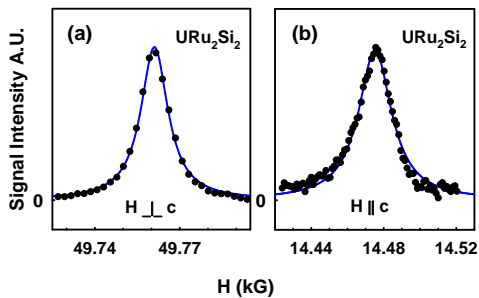}}
\caption{$^{29}$Si~NMR spectra in URu$_2$Si$_2$ for $T=14.5$~K 
under (a) transverse- and (b) longitudinal-field conditions.
Curves: fits to Lorentzian functions of HWHM $\Gamma({\bf H},T)$.
}
\label{spc}
\end{figure}

Spectra for ${\bf H}\perp{\bf\hat{c}}$ and ${\bf H}\parallel{\bf\hat{c}}$ 
[Fig.~\ref{spc} (a) and (b) respectively] consisted of a single narrow line.
Each line could be fit to a Lorentzian function of 
half-width-at-half-maximum (HWHM) $\Gamma$
for the $T$- and ${\bf H}$-range reported here.
We find that the linewidth can be written as
$\Gamma^2({\bf H},T)=\Gamma_{m}^2({\bf H},T)+\lambda^2(T)$; 
where $\Gamma_{m}$ is the contribution due to the sample 
magnetization (a term proportional to $H$), and $\lambda$ is the 
new contribution to the width, which is zero above~$T_N$.

\begin{figure}[h]
\centerline{\epsfxsize=2.5in \epsfbox{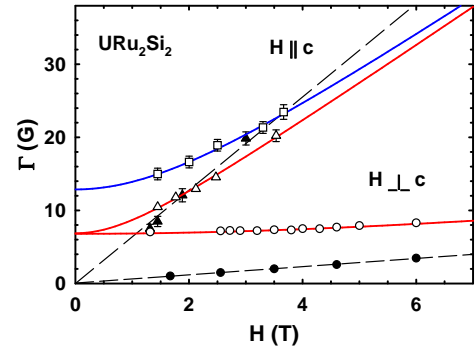}}
\caption{Field dependence of the full $^{29}$Si linewidth $\Gamma$.
Circles: ${\bf H}\perp{\bf\hat{c}}$, $T=14.5$~K (open) and
$T=20$~K (filled). Filled triangles: ${\bf H}\parallel{\bf \hat{c}}$,
$T=20$~K. Open triangles: ${\bf H}\parallel{\bf\hat{c}}$, 
$T=14.5$~K. Squares: $T=$4.2~K.
Curves: one- (dashed lines) and two- (solid) component fits to the linewidth.}
\label{field}
\end{figure}
Separating $\Gamma$ into two components is compelled by its field 
dependence, which we present in Fig.~\ref{field}.
For ${\bf H}\perp{\bf\hat{c}}$ and $T\sim{14.5}$~K, $\Gamma$ (open circles) 
is weakly linear in the field magnitude above about 3--4~T and field 
independent below this value, with a clear non-zero intercept.
The data can be fit to the assumed form as indicated by the curve
[i.e., $\lambda(14.5$~K$)\sim{7}$~G].
For comparison, we present a similar study at $T=20$~K (solid circles),
where no extra width exists: the dashed curve is a straight-line fit
(slope$\sim$0.6~G/T).
Similarly, the solid triangles in Fig.~\ref{field} represent the total 
width for ${\bf H}\parallel{\bf\hat{c}}$ and $T=20$~K.
Here, the slope ($\sim{6.4}$~G/T) [dashed-line fit]  is more than ten times
greater than for ${\bf H}\perp{\bf\hat{c}}$ due to the large magnetic 
anisotropy of the system.
The open triangles and the squares are obtained for $T=14.5$, and 4.2~K
respectively.
As for ${\bf H}\perp{\bf\hat{c}}$, the curves drawn through these data 
are fits to a two-component $\Gamma$.
It is seen from these fits, that the orientation dependence of 
$\lambda$ at constant $T$ is weak or nonexistant; for example,
at 14.5~K the fits are consistent with having similar intercepts, 
i.e., 
$\lambda({\bf H}\perp{\bf\hat{c}},14.5$~K$){\sim}\lambda({\bf H}
\parallel{\bf\hat{c}},14.5$~K$)\sim$~7~G.
[We have also measured the orientation dependence of $\lambda$ directly;
the results are discussed below.]
The value of $\lambda$ at 4.2~K is clearly different, 
$\lambda(4.2$K$)\sim{13}$(2)~G, as one could expect for a $T$-dependent 
$\lambda$ (see Fig.~\ref{lambda} below).
Finally, one can see from Fig.~\ref{field} why, in early experiments, 
$\lambda$ was not detected with NMR at 
high-${\bf H}\parallel{\bf\hat{c}}$ fields~\cite{bernaljapan} 
or for poorly aligned samples~\cite{kohori}.
Because of the strong magnetic anisotropy in this system, the part of 
the width due to either the magnetization in the aligned powder, or 
the anisotropic residual powder pattern in poorly aligned samples, 
can overwhelm $\lambda$, even at low values of ${\bf H}\parallel{\bf\hat{c}}$.

\begin{figure}[h]
\centerline{\epsfxsize=2.5in \epsfbox{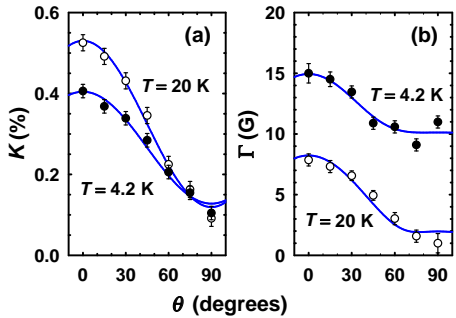}}
\caption{Dependence of $^{29}$Si Knight shift $K$ (a) and 
linewidth $\Gamma$ (b) on {\em c}-axis angle $\theta$ in URu$_2$Si$_2$ 
(see text).}
\label{theta}
\end{figure}

The Knight shift $K$ has been reported previously under longitudinal 
and transverse field geometries~\cite{kohori,kohori2k,bernaljapan}.
Since in antiferromagnetic systems one might expect a change in the 
orientation dependence of $K$ as $T$ crosses $T_N$~\cite{jaccarino},  
we followed $K$ vs. $\theta$, the field orientation angle, above 
and below $T_N$.
We find that the behavior of $K$ reflects only the anisotropic magnetization
and does not seem to be involved in the linewidth effect we observe: the 
lineshape (Lorentzian-like) does not change with $\theta$ 
(Fig.~\ref{spc}) and does not shift anomalously through $T_N$.
This can be seen clearly in Fig.~\ref{theta}~(a) and (b) where we 
present, respectively, $K(\theta)$ and $\Gamma(\theta)$ above 
($T=20$~K; open circles) and below ($T=4.2$~K; closed circles) $T_N$.
For both temperatures, the shift can be fit to 
$K=\mu^2{K_{\|}}(T)+(1-\mu^2)K_{\bot}$, $\mu\equiv\cos\theta$ (drawn lines), 
as expected from crystal anisotropy~\cite{nmr}.
The difference in magnitude for $\theta\rightarrow$~0 is due to the 
temperature dependence of the magnetization in that direction.

The curve drawn in Fig.~\ref{theta}(b) for $T=20$~K is expected 
if $\Gamma$ represents a distribution of anisotropic shifts with two 
independent width components: 
$\delta{K_\|}(T)$ and $\delta{K_\bot}$ 
(i.e., $\mu$ is not distributed, and $\delta{K_\bot}$ is not
$T$-dependent).
We find $\delta{K_\|}(20$~K$)=0.056(4)$~\%, and 
$\delta{K_\bot}=0.013(4)$~\%.   
For $T=4.2$~K [Fig.~\ref{theta} (b), closed circles], 
$\lambda$ is present, and the fit is consistent with 
$\lambda$ being isotropic.
Here, we find $\delta{K_\|}(4.2$~K$)=$0.078(4)~\%, and
$\lambda(4.2$~K$)=$9.9(5)~G. 

Figure~\ref{lambda} gives the temperature dependence of $\lambda$.
The filled symbols are obtained in the transverse-field geometry
by direct quadrature subtraction of the term 
$\Gamma_m({\bf H}\perp{\bf\hat{c}})$, which is temperature independent.
Since $\Gamma_m$ is strongly $T$-dependent in the longitudinal field
configuration, direct extraction of $\lambda$ from the temperature dependence 
of the total width $\Gamma$ in that geometry is not straightforward.
Our procedure is as follows.
We subtracted $\Gamma_m({\bf H}\parallel{\bf\hat{c}},T)$ in this case by using 
the data in 
Fig.~\ref{field} together with a scaling of the form
$\Gamma_m({\bf H}\parallel{\bf\hat{c}},T)\propto{M({\bf H}\parallel{\bf\hat{c}},T)}$, 
where $M$ is the magnetization.
Random errors in the values of ${\bf H}\parallel{\bf\hat{c}}$, 
$\theta$ and $T$ from different
experimental runs become important in deducing a good value of 
$\lambda(T)$ as can be seen from the scatter of the data in this 
geometry.
Nevertheless, the resulting values are consistent with the lack
of anisotropy inferred above from field and orientation studies.
The behavior of $\lambda$ clearly signals that there exists a
coupling between the $^{29}$Si nuclear spins and the electronic 
transition at $T_N$.
This is further corroborated by the BCS-gap-equation fit (drawn curve 
in Fig.~\ref{lambda}), which indicates a mean-field-theory 
temperature dependence for $\lambda$.
\begin{figure}[h]
\centerline{\epsfxsize=2.0in \epsfbox{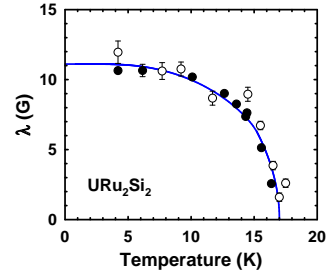}}
\caption{Anomalous ${\lambda}(T)$ extracted from $^{29}$Si NMR linewidths
in URu$_2$Si$_2$ for ${\bf H}\perp{\bf\hat{c}}$ (filled circles) and 
${\bf H}\parallel{\bf\hat{c}}$ (open circles).
Curve: fit to BCS-gap equation.}
\label{lambda}
\end{figure}

Of the characteristics found for $\lambda$ during this study,
its lack of strong anisotropy and its independence of the lineshift $K$
are the hardest ones to reconcile within a simple model of magnetic 
order/disorder at the U sites as the origin of $\lambda$.
Although a distribution of internal fields could be produced by disorder 
or incommensuration in the direct (dipolar) or indirect transferred 
interaction between U~moments and the $^{29}$Si nuclei, one would expect 
the linewidth to have the anisotropy of this interaction and the lineshape 
and width to change as a function of the external field direction.
Neither of these expectations is borne out by the behavior of $\lambda$,
even though the shift $K$ does follow the anisotropic susceptibility.
Accordingly, the observed lineshape
(Fig.~\ref{spc}) suggests that a fair number of nuclei,
contributing to the center of the line, probe little or no internal field
independently of the external-field orientation.
Any model of magnetic disorder at the U-sites, 
would have to invoke a continuous distribution of moment sizes or domains 
that includes values all the way down to zero.
Although domains of this type are found in, for example, NiO, as 
reported 40 years ago~\cite{roth}, the direction of the spontaneous  
moment within the domains would have to be either distributed almost 
isotropically (for $\mu\sim{0.03}\mu_B$, 0$^\circ\lesssim\theta\lesssim 80^\circ$),
or such that it is coupled to the direction of the applied field.
In any case, such interpretation would be in striking contrast with ND
and $\mu$SR measurements for which the correlations lie predominantly along 
the $c$~axis.
Thus, the behavior of $K$ and $\lambda$ together with ND and $\mu$SR results
rule out static magnetic order/disorder at the U sites.

Alternatively, a qualitative explanation of the main characteristics of 
$\lambda$ is achieved by invoking an indirect nuclear spin/spin interaction 
between unlike nuclei: ${\cal H}_{ij}=A_{ij}{\bf I_i \cdot I_j}$.
For instance, $\lambda$ is static and inhomogeneous, so it represents a 
distribution of time-average-effective local fields such as could be produced 
by $^{99,101}$Ru nuclei at $^{29}$Si sites.
The disorder effect would come from the fact that all isotopes involved are 
randomly distributed in the crystal because of their low natural abundance. 
This mechanism does not contribute to the line shift $K$, which explains the
independence of $\lambda$ with respect to this parameter.
The field independence of $\lambda$ could also be explained by this mechanism; 
the indirect interaction is a second order effect, dominated by the excitation 
energy of the mediating electrons, which is in general much larger than the 
nuclear resonance energy.

The difficulty here is to ascertain what the unperturbed Hamiltonian ought 
to be in order to calculate $A_{ij}$ from first principles.
For example, RKKY interactions between the nuclei~\cite{nmr,bloembergen}
could be mediated either by normal conduction electrons or by renormalized 
quasiparticles.
Crude estimates based on our previous measurements of the Knight shift vs. 
the spin susceptibility~\cite{bernaljapan} yield $A_{ij}\sim 10^{-6}$~G 
in the former case, and $A_{ij}\sim 10^{-3}$~G in the latter.
These values are negligible and would also be present in 
the high-temperature state.
The possibility that such an effect could be amplified to $\sim$10~G by 
the phase transition would require an increase of order 10$^2$ in the overlap 
of the quasiparticle wavefunctions with the $^{29}$Si~nuclei below~$T_N$.
Another possibility is the Suhl-Nakamura mechanism~\cite{nakamura}, in 
which antiferromagnetic spin waves mediate the indirect nuclear spin-spin  
interaction.
However, in view of the fluctuating-moment picture (which is supported by our 
measurements) \cite{kohori2k}, the use of a static AF state for the electron 
system is not justified.

On the other hand, the evidence that the primary hidden order parameter~$\psi$
(whatever it might be) is actually coupled to the antiferromagnetic correlations
observed by ND and $\mu$SR allows one to envision the following mechanism for 
indirect nuclear spin interactions.
Let a $^{99,101}$Ru~nucleus produce a virtual excitation of the electronic state 
(governed by~$\psi$).
Because of the coupling between $\psi$ and the secondary (antiferromagnetic) 
order parameter $m$~\cite{mydosh2k}, a change in the electronic
spin state occurs allowing the nearby $^{29}$Si~nuclei to detect the interaction. 
For instance, if the primary order parameter is quadrupolar or that of a 
charge-density wave CDW, there will
be an interaction between the $^{99,101}$Ru~nuclei and the electric
field gradients produced by the charge state of the U~ions.
The presence of a small external field (small compared with the excitation 
energy of the quadrupolar states of the electrons) produces a Zeeman 
splitting of the electron spin states, so that electron spin flips can 
occur when a quadrupolar/CDW excitation is induced by the $^{99,101}$Ru~nuclei.
The time average of these spin-flips would be responsible for the local
change in resonance frequency at the $^{29}$Si sites.
Unfortunately, we lack at present a good characterization of the electronic 
ground state to proceed further in this direction.
Nevertheless, we suggest that our conjecture be checked theoretically 
using the different microscopic models that have been proposed in the 
literature.

In conclusion, magnetic U-site order/disorder as the origin of the anomalous
linewidth $\lambda$ can be ruled out by combined NMR and ND/$\mu$SR experiments.
Furthermore, we discovered a relationship between $^{29}$Si~NMR linewidth 
and the primary ``hidden'' order parameter of the~17~K transition.
Double resonance experiments in which the nuclei are decoupled by an rf field
and NMR in isotopically enriched samples need to be performed as functions of 
$T$ and ${\bf H}$ in order to elucidate the extent to which indirect coupling 
between unlike nuclei is the source of the NMR broadening.
Finally, we point out that we appear to be observing the effects of charge 
degrees of freedom in URu$_2$Si$_2$ by means of $^{29}$Si, which has a spin 
1/2 and cannot sample quadrupolar effects directly.
The proposed experiments should further test this conjecture.

This work was supported by NSF grants DMR-9820631 (CSULA), DMR-9731361 (UCR),
by awards from the Research Corporation (CSULA and Whittier), and by the
Dutch Foundation FOM (Leiden University).


\begin{references}
\bibitem{comment} We first reported a linewidth increase below
$T_N$ for a field in the ($a,b$) plane in Ref.~\cite{bernaljapan}.
\end{references}
\end{document}